\documentclass{ws-rv9x6}

\usepackage{graphicx}
\usepackage{latexsym}

\begin{document}

\setcounter{chapter}{0}

\chapter{UNDER-KNOTTED AND OVER-KNOTTED POLYMERS: 1. UNRESTRICTED LOOPS}

\markboth{N.T. Moore, R.C. Lua, A.Y. Grosberg}{Under-Knotted and
Over-Knotted Polymers: 1.  Unrestricted Loops  }

\author{Nathan T. Moore, Rhonald C. Lua, Alexander Yu. Grosberg}

\address{Department of Physics, University of Minnesota,Minneapolis, MN 55455, USA}

\begin{abstract}
We present computer simulations to examine probability
distributions of gyration radius for the no-thickness closed
polymers of $N$ straight segments of equal length.  We are
particularly interested in the \emph{conditional} distributions
when the topology of the loop is quenched to be a certain knot
${\cal K}$.  The dependence of probability distribution on length,
$N$, as well as topological state ${\cal K}$ are the primary
parameters of interest. Our results confirm that the mean square
average gyration radius for trivial knots scales with $N$ in the
same way as for self-avoiding walks, where the cross-over length
to this "under-knotted" regime is the same as the characteristic
length of random knotting, $N_0$. Probability distributions of
gyration radii are somewhat more narrow for topologically
restricted under-knotted loops compared to phantom loops, meaning
knots are entropically more rigid than phantom polymers.  We also
found evidence that probability distributions approach a universal
shape at $N>N_0$ for all simple knots.
\end{abstract}

\section{Introduction}

\subsection{The goal of this work}

Consider a random closed polygon of some $N$ segments, all of
equal length $\ell$.  What is the probability $w_{\rm triv}(N)$
that this polygon, considered as a closed curve embedded in $3D$,
is topologically equivalent to a circle, that is, represents a
trivial knot?  What is the probability $w_{\cal K}(N)$ that it
represents a knot of any other kind, ${\cal K}$? Such questions
arose first in the context of DNA\cite{old1} and other
polymers\cite{old2}, and continue to attract significant attention
to the present day. Although a large body of information has
accumulated, mostly through computer
simulations\cite{Maxim,koniaris_muthu_N0,deguchi_N0}, final
mathematical understanding of these questions remains elusive,
despite their elementary formulation.

Meanwhile, a new set of questions came to the forefront in the
last several years.  For instance, what is the \emph{conditional}
probability density of the loop gyration radius given that its
topology is fixed to be ${\cal K}$?  As a first step, what is
the average gyration radius of the loop with the given knot state
${\cal K}$?  This latter question was first discussed by des
Cloizeaux\cite{conj1} and then re-visited
theoretically\cite{Quake,AG_pred} and
computationally\cite{Deutsch_support,swiss_PNAS,Degichi_excluded_volume_limit,Deguchi_finite_size,Deguchi_one_more,RG_style_fitting,Nathan_PNAS}.
The excitement in the field is partially driven by the idea, first
conjectured in the work\cite{conj1}, that topological constraints
act effectively like self-avoidance, leading to the non-trivial
scaling $\langle R_g^2 \rangle \sim N^{2 \nu}$, where $\nu$ is the
critical exponent known in the theory of self-avoiding walks, $\nu
\approx 0.588 \approx 3/5$.

The distinction between the two groups of questions can be
illuminated by the comparison with the concepts of annealed and
quenched disorder, well known in the physics of disordered systems
(see, for instance, book\cite{Ziman_disorder_book}).  If the loop
is phantom, i.e. if it can freely cross itself, then its
topological state is annealed.  In this case, we can ask what the
probability is to observe a certain topological state ${\cal K}$.
For the loop which is not phantom and cannot cross itself, the
knot state is frozen, or quenched, and we can discuss physical
properties of the loop, such as its size or entropy for every
given knot state ${\cal K}$.

The main goal of this paper is to look more closely at the
probability distributions of the gyration radius of the loops
which are topologically constrained but not constrained otherwise.
In section \ref{sec:over}, we provide an overview of the previous
results about the mean square averaged gyration radius as well as
some related questions of method and simulation technique.  We
shall concentrate on the relatively simple knots, such as $0_1$,
$3_1$, and $4_1$, formed by rather long polymers, with $N$ up to
$3000$.  Using the terminology introduced in the recent work
\cite{Nathan_PNAS}, we can say we shall be interested mostly in
the under-knotted regime. This terminology makes simultaneous use
of both annealed and quenched views of polymer topology.  The idea
is as follows.  Consider real polymer loop with some quenched knot
${\cal K}$.  It is considered over-knotted if upon topological
annealing, allowing loop states to be sampled without topological
constraints, the loop is likely to become topologically simpler
than ${\cal K}$. Otherwise, the loop is considered under-knotted.
Roughly, loop is under-knotted if it "wants" to have more knots,
and it is over-knotted if it "wants" to have fewer knots.  Whether
a quenched loop is over- or under-knotted depends on the number of
segments, $N$, and, in general, on some other conditions, such as
solvent quality and the like.  It is because the loop is
under-knotted that it may swell, even if there is no excluded
volume or self-avoidance.  Here, however, terminology
clarification is in order.

\subsection{Some terminology: non-phantom polymers and
self-avoiding polymers are two different things}

We should first emphasize the difference between concepts of
self-avoiding polymers and non-phantom polymers.  These two
concepts are quite frequently confused.  The idea of
self-avoidance always involves certain finite non-zero length
scale, let say $d$, such that two pieces of a polymer cannot
approach each other closer than $d$. For instance, if one thinks
of a polymer as a little garden hose, then $d$ is its diameter.
Real polymers, of course, always have some excluded volume, or
some thickness $d$.  On the other hand, polymers which we call
phantom are imagined to be able to switch from an under-pass to
over-pass conformations, but, importantly, neither former nor
later state violate the self-avoidance, or excluded volume,
condition.  Speaking about phantom polymers, we should
intentionally close our eyes on the process - \emph{how} the
polymer passes from under- to over- state.  This question is
irrelevant when we address probabilities or equilibrium
statistical mechanics.  In some sense, the idea of a phantom
polymer can be illustrated by the properties of a DNA double helix
in the presence of topo-II enzymes\cite{topo-II}.  Of course, this
question of crossing mechanism becomes decisive if one wants to
look at polymer dynamics without enzymes; for the studies of
dynamics, the phantom model is meaningless, one should think in
terms of reptation instead\cite{DoiEdwards}.

On a more quantitative level, it is known for the polymer with $N$
segments of the length $\ell$ and diameter $d$ that the excluded
volume effect does not lead to appreciable swelling as long as $N
\ll (\ell /d)^{2}$ (see, e.g., book\cite{AG_Red}, page 91).  For
dsDNA at a reasonable ionic strength, this implies chain length up
to about $2500$ segments, or $75000$ base pairs.  In this sense,
our testing of loops up to $N=3000$, although dictated by our
computational possibilities, is also meaningful for the important
particular case of DNA.  On this length scale, it is quite
reasonable to neglect the self-avoidance condition, and at the same
time to work with the polymer which is not phantom, because its
topological state is quenched (unless enzymes are present).

\section{Brief overview of our recent work\protect\cite{Nathan_PNAS}}\label{sec:over}

Our most recent work has investigated the average size of knotted
loops.  The initial focus was on those loops with trivial knot
topology, denoted $0_1$, as their size has been addressed
theoretically \cite{conj1,AG_pred}.  In collecting data through
simulation we were able to gather statistically significant
information about several other knots of low prime crossing
number, specifically, $3_1$, $4_1$, $5_1$ and $5_2$ knots.

\subsection{Simulation methods}

Like others\cite{Deguchi_one_more,swiss_PNAS,RG_style_fitting}
our initial approach to the problem was to generate loops, compute
the gyration radius for each of them
\begin{equation} R_g^{2} = \frac{1}{2N^{2}} \sum _{i=1}^{N} \sum_{j=1}^{N}
r_{ij}^{2} \ , \ \  r_{ij} = \left| {\vec r}_i - {\vec r}_j
\right| \ , \label{eq:gyration}  \end{equation}
${\vec r}_i$ being the position vector of the joint number $i$,
and then analyze the generated conformations with several
topological invariants.

Our loop generation routine is discussed in the
\ref{sec:Loop_Generation}.  Importantly, the loops are generated
without any relation towards their topology.  When a loop is
generated, its knot type is assigned.  Therefore, we can use the
ensemble of all generated loops to address questions regarding the
annealed topology, such as the population fractions of various
knots.  At the same time, we are able to determine average size,
and more generally, the probability distribution of size for loops
assigned any given knot type, which means, we can address the
quenched topology questions.

To determine loop topology we compute several topological
invariants\cite{the_knot_book}.  For the loops with $N \le 300$,
we used Alexander invariant $\Delta (-1)$ and Vassiliev invariants
of degree $2$ and $3$, $v_2$ and $v_3$. The loop was identified as
a trivial knot when it yielded $ \left|\Delta (-1) \right| = 1$,
$v_2 = 0$, and $v_3=0$.  For longer loops of $N>300$, we were able
to use only $\Delta (-1)$ and $v_2$ invariants, assigning trivial
knot status to the loops with $ \left|\Delta (-1) \right| = 1$,
and $v_2 = 0$. The details of our computational implementation of
these invariants are described
elsewhere\cite{Rhonald_implementation}. Of course, because of the
incomplete nature of topological invariants, our knot assignment
is only an approximation, and surely was sometimes in error.

\subsection{Knot population fractions}

We begin by addressing the annealed topology questions.

Theoretically, it is believed that the probability of a trivial
knot is exponential in $N$:
\begin{equation} w_{\rm triv} (N) = w_{0} \exp \left( - N / N_0 \right) \ ,
\label{eq:probab}  \end{equation}
at least, asymptotically when $N \gg 1$.  Such exponential
behavior was observed in a number of simulation works for a
variety of models \cite{koniaris_muthu_N0,deguchi_N0}.  By now, it
is already considered "obvious" by physicists in the field.  It is
indeed fairly obvious through the intuition gained by the
study\cite{Spitzer,Edwards,Prager_Frisch} (see also more recent
work\cite{Bessel} and references therein)  of exactly solvable
model of winding around a point or a disc in $2D$.  This model
shows that typical Brownian trajectory (that is, polymer with
$\ell \to 0$ and $N \to \infty$) tends to produce a diverging
winding angle, that is, an infinite number of turns around the
point-like obstacle.  It does not seem to require a particularly
great leap of imagination to conclude that at very large $N$ some
finite scale knots should be formed with a non-zero frequency
everywhere along the polymer - and this exactly leads to
Poisson-like exponential formula (\ref{eq:probab}).

With regard to the probabilities of other non-trivial knots, it may be
argued that they should also be asymptotically exponential
\begin{equation} w_{\cal K} (N) = w_{0}^{(\cal K)} \exp \left( - N /
N_{\cal K} \right) \ , \label{eq:probabK}  \end{equation}
and, moreover, that characteristic length should be the same as
that for trivial knots: $N_{\cal K} = N_{0}$.  This latter idea
can be understood by saying that for every knot, the loop must
eventually become strongly under-knotted if $N$ increases without
bound while knot is quenched.  Formula (\ref{eq:probabK}) was also
tested, albeit by a smaller number of
simulations\cite{deguchi_N0,Deguchi_lectures_at_knots}.

In the work\cite{Nathan_PNAS}, we fit formula (\ref{eq:probab}) to
our trivial knot data and found critical length, $N_0=241 \pm 0.6$
and $w_0 = 1.07 \pm 0.01$.  This value of $N_0$ is consistent with
the result on a rod-bead
model\cite{koniaris_muthu_N0,Degichi_excluded_volume_limit} in the
limit of excluded volume radius sent to zero.  In other
works\cite{koniaris_muthu_N0,deguchi_N0} somewhat larger values of
$N_0$ were reported, closer to $300$ or $330$.  We interpret this
discrepancy as being due to the fact that we examined the model
with all segments of the same length, while the
works\cite{koniaris_muthu_N0,deguchi_N0} dealt with Gaussian
distributed segments.  We consider it an exciting challenge to
understand why these two models exhibit differing values of
characteristic knotting length.

\begin{figure}
\centerline{\scalebox{0.45}{
\includegraphics{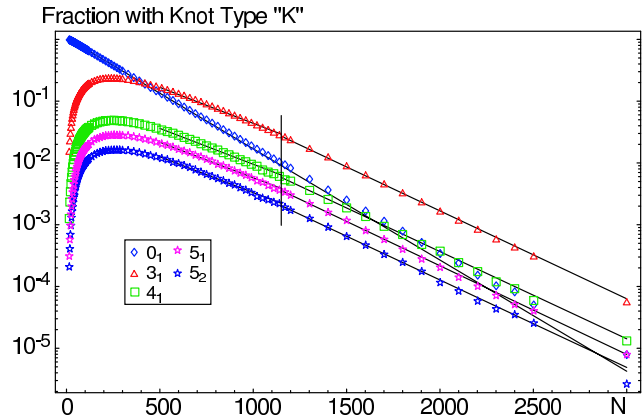}}}
\caption{The fraction of loops generated with trivially knotted
topology followed the well known exponential form, equation
(\protect\ref{eq:probab}), as a function of loop length $N$.
Deviation from the fit line at large $N$ is due to the
incompleteness of topological invariants employed and reflects
contamination of the supposedly trivial pool with some non-trivial
knots.  The fractional population curves for several different
simple knot types are shown and labeled.  Although their overall
decay can be reasonably fit by exponents, the characteristic
lengths $N_{\cal K}$ appear larger than $N_0$, which probably
means that true asymptotics are very slowly achieved.}
\label{fig:fraction}
\end{figure}

Figure \ref{fig:fraction} shows our simulation data for trivial
knots fraction, along with the data illustrating the relative
frequency of other knot types.  To the accuracy of our
simulations, we do not see all non-trivial knot probabilities
decaying with the same characteristic length $N_0$.  However, we
tried to determine $N_{\cal K}$ (see equation (\ref{eq:probabK}))
by fitting the data over sliding window.  For instance, Table
\ref{tab:multiple_fraction_fit} shows the fit parameters obtained
on the interval $500 < N < 1150$, or on the interval on $1150 < N
<3000$.  It is clearly seen that "apparent" characteristic length
decreases.  Although far from proof, this result is consistent
with the theoretical argument behind formula (\ref{eq:probabK})
and allows one to hypothesize that the asymptotics is just very
slowly achieved.

\begin{table}
\tbl{Characteristic Lengths, $N_{\cal K}$}
{\begin{tabular}{|c|c|c|} \hline
knot type & $N_K$ on   & $N_K$ on \\
& $(500<N<1150)$ & $(1150<N<3000)$ \\
\hline
$0_1$ & $241$ & $250$ \\
$3_1$ & $373$ & $305$ \\
$4_1$ & $374$ & $307$ \\
$5_1$ & $375$ & $307$ \\
$5_2$ & $378$ & $302$ \\
\hline
\end{tabular}}
\label{tab:multiple_fraction_fit}
\end{table}

\subsection{Average size of different knots}

\subsubsection{Scaling of the trivial knot size}

When averaged over all loops, the mean square gyration radius,
$\langle R^2_g \rangle$, is equal to $N \ell^2 /12$, which is two
times smaller than the similar quantity for linear chains (see,
for instance, book\cite{AG_Red}; see also
\ref{sec:Rg_distribution}).  As regards $\langle R^2_g \rangle$
averaged over only trivially knotted loops, the
theorists\cite{conj1,Quake,AG_pred} predicted, that trivial knots
develop swelling behavior for $N \gg N_0$, in a way similar to
objects which experience excluded volume forces:
\begin{equation} \langle R_g^{2} \rangle_{\rm triv} = \left\{
\begin{array}{lcr} \left( \ell^{2} /12 \right) N & {\rm if} & N
\ll N_0 \\ A \left( \ell^{2} /12 \right)  N^{2 \nu} & {\rm if} & N
\gg N_0
\end{array} \right. \ , \label{eq:predict}  \end{equation}
where scaling power is $\nu \approx 0.589$, and where $N_0$ is the
same parameter introduced in formula (\ref{eq:probab}).

We want to emphasize here that the first line of the prediction,
formula (\ref{eq:predict}), is not connected to any delicate and
thus possibly unreliable theoretical arguments, but rather comes
out of almost pure common sense.  Indeed, when $N \ll N_0$,
according to formula (\ref{eq:probab}), there is only marginal
probability for a phantom loop to have any knot other than
trivial.  This means, the ensemble of trivially knotted loops at
these $N$ very nearly coincides with the ensemble of all loops,
for which $\langle R_g^2 \rangle$ is certainly equal to $N \ell^2
/12$.

\begin{figure}
\centerline{\scalebox{0.45}{
\includegraphics{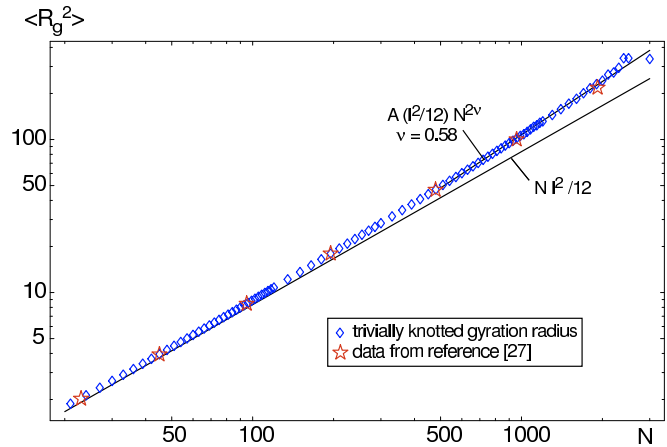}}}
\caption{Gyration radius averages over trivially knotted loops.
The trivial knot average exhibits power law behavior at large $N$
similar to that experienced by polymers which have excluded
volume.  The trivial knot data is systematically larger than the
average over all loops, shown as the solid line in the figure.
This topology driven swelling is seen to develop beyond the
critical length about $N_0=241$.  Independently collected
data\protect\cite{VOLOGODSKII_private} is shown by stars ($\star$)
and agrees with our results. } \label{fig:radius}
\end{figure}

Figure (\ref{fig:radius}) demonstrates how our simulation results
are consistent with formula (\ref{eq:predict}).  First of all, we
see that indeed the Gaussian scaling $\langle R_g^2 \rangle = N
\ell^2 /12$ is recovered at $N$ below $N_0$.  Fitting the data
over the interval $500 < N < 2500$, we found the parameters, $\nu
\approx 0.58 \pm 0.02$ and $A \approx 0.44 \pm 0.03$.  It is not
only important that $\nu$ is consistent with expectations, it is
also important that the value of pre-factor $A$ provides for
smooth cross-over between regimes at $N$ very close to $N_0$, as
expected (because $A$ is very close to $N_0^{1-2\nu} \approx
0.42$).

\subsubsection{Corrections to scaling}

Can one pull the analogy between trivial knots and self-avoiding
polymers further?  The temptation in the field
\cite{swiss_PNAS,Deguchi_one_more,Deguchi_finite_size} has been to
fit the trivial knot data with a more complex perturbation
formula, motivated by the analogy with the excluded volume
problem\cite{RG_style_fitting},
\begin{equation} \langle R_g^{2} \rangle = A \frac{\ell^2}{12} N^{2 \nu} \left[ 1 + B \left(
\frac{N_{0} }{ N} \right)^{ \Delta} + C \left( \frac{N_{0} }{ N}
\right)^{ 2 \Delta} + \ldots \right] \ . \label{eq:interpolation}
\end{equation}

To understand this formula, it is useful to recall its appearance
in the better known context of the excluded volume problem (here,
we re-phrase presentation in the book\cite{AG_Red}).  For the
excluded volume (or self-avoiding) polymer, one first shows that
gyration radius can be written in the form $\langle R_g^2 \rangle
= N \ell^2 f(x)$, where $f(x)$ is a universal function of the
argument $x = (d/\ell) \sqrt{N}$ (where $d$ and $\ell$ are segment
thickness and length, respectively).  For our purposes here, we
denote $N^{\star}_0 = (\ell / d)^2>1$ and then write $x= \left(
N/N^{\star}_0 \right)^{1/2}$. When $x$ is small, $x \ll 1$, then
$f(x)$ can be presented as an (asymptotic) perturbation series in
integer powers of $x$. When $x$ is large, $x \gg 1$, the leading
term in $f(x)$ contains the non-trivial scaling power: $f(x) \sim
x^{2 \nu -1}$, and then the correction terms in this large $x$
asymptotics involve negative powers of $x$, in most cases
believed\cite{RG_style_fitting} to be integer negative powers:
$f(x) \sim x^{2 \nu -1} \left[1 + B/x + C / x^2 + \ldots \right]$.
Using this formula to write $\langle R_g^2 \rangle$ in terms of
$N_0^{\star}$, we obtain exactly the equation
(\ref{eq:interpolation}) (with $\Delta = 1/2$).

This consideration shows that for the excluded volume problem
formula (\ref{eq:interpolation}) is only valid at $x \gg 1$, or $N
\gg N_0^{\star}$.  It is \emph{not} an interpolation formula valid
across the cross-over region $x \sim 1$; it does not connect two
asymptotics smoothly.  For the latter reason, it cannot be
considered an interpolation for trivial knots.  That is why we
think it is not correct to fit the simulation data to this formula
in the range of $N$ other than $N \gg N_0$ (or at least $N >
N_0$).

Unfortunately, our data do not allow for reasonable fit to this
formula even in the range $N > N_0$.  The reason is seen in the
fact that our data represent a curve which seems to keep bending
upwards as $N$ increases, while formula (\ref{eq:interpolation})
implies saturation of the log-log slope to that dictated by the
power $2 \nu$.  A mechanical attempt to fit the formula to the
data yields physically meaningless values for $\nu$ which are
greater than unity.

Currently we do not know why data do not fit formula
(\ref{eq:interpolation}).  One reason may be simply poor
statistics and noisy character of data at large $N$.  It might
also be an indication of the knot pool contamination at large $N$
because of the incompleteness of topological invariants. This is
possible, but, in our opinion, not very likely given that trivial
knot fraction does not deviate much from the exponential fit (see
Figure \ref{fig:fraction}).  In the work\cite{Nathan_PNAS}, we
attempted to address this question deeper, introducing the
correction for the errors in knot assignment.  It did not yield
much change in terms of $\langle R_g^2 \rangle$, making us a bit
more confident that the problem might be somewhere else.  For
instance, it is possible that the formula (\ref{eq:interpolation})
does not apply to trivial knots, indicating some restricted
applicability of the very analogy between trivial knots and
excluded volume polymers. Much work will be necessary to clarify
this issue.

\subsubsection{Averaged sizes of non-trivial knots}

\begin{figure}
\centerline{\scalebox{0.45}{
\includegraphics{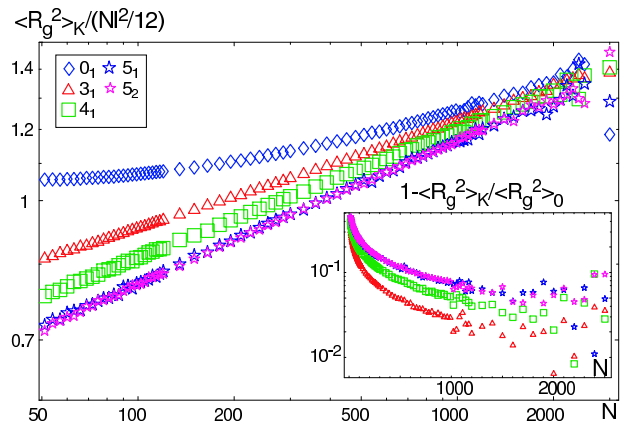}}}
\caption{Log-log plot of the mean square gyration radius, $\langle
R_g^{2} \rangle_{\cal K}$, of knot type $\cal K$, normalized by
the topology blind average over all loops for several particular
knot types. The inset, which shows the ratio of a particular knot
gyration radius to the trivial knot gyration radius, $1 - \langle
R_g^2 \rangle_{\cal K}/\langle R_g^2 \rangle_{0}$, demonstrates
that all knots remain smaller than, but approach the size of,
trivial knots. } \label{fig:topology_zoo}
\end{figure}

Our measurement of the swelling of non-trivial knots is shown in
figure (\ref{fig:topology_zoo}).  It is overall consistent with
findings by earlier works\cite{swiss_PNAS,Deguchi_one_more}.  We
find that the simple knots cross over from an over-knotted state,
in which they are much smaller than the average sized loop to an
under-knotted state in which they seem to approach the scaling of
trivial knots in an asymptotic fashion. The inset in this image
shows this asymptotic approach in the form of a small parameter,
$\beta = 1 - \langle R_g^2 \rangle_{\cal K}/\langle R_g^2
\rangle_{0}$ decaying with increasing $N$.

\section{Probability distributions of the loop
sizes}\label{sec:probabilities}

Our data allow us to make one more step and to look not only at
the averaged value of $R_g^{2}$ for trivial and some non-trivial
knots, but also at the entire probability distributions. We were
able to generate and analyze histograms of quality (i.e. looking
smooth when plotted, a minimum of $10^5$ loops for 
each curve) for loops of size $N \le 1200$. Predictably, the probability
distributions are different for different topological classes,
such as all loops versus loops of a certain knot type ${\cal K}$.
Also predictably, the probability distributions of $R_g^{2}$
spread out as $N$ increases.  The latter observation suggests the
idea of looking at the probability distributions of the re-scaled
variable $\rho = R_g^{2}/\langle R_g^{2} \rangle$, where the
normalization factor $\langle R_g^{2} \rangle$ is taken separately
for each $N$ and for each topological entity.

\begin{figure}
\centerline{\scalebox{0.45}{
\includegraphics{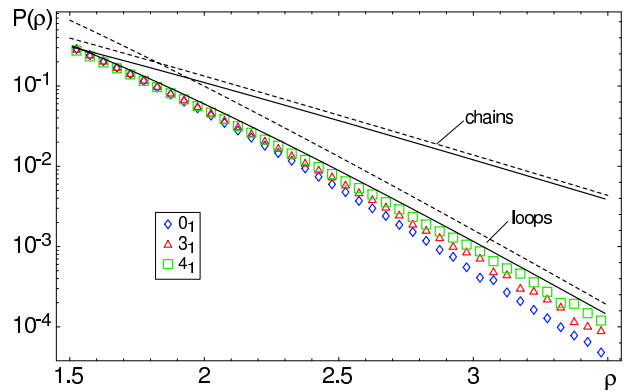}}}
\caption{The probability density plot for
chains\protect\cite{Fixman} (line), all loops (another line), and
loops with certain knots ($0_1$ - $\diamond$, $3_1$ - $\Delta$,
$4_1$ -$\Box$) in the range of large $\rho > 1$. Distributions are
presented in terms of the scaling variable $\rho = R_{g}^{2} /
\langle R_{g}^{2} \rangle$. The asymptotics calculated in \ref{sec:Rg_distribution}, 
equations (\ref{eq:asymp_chain}) and (\ref{eq:asymp_loops}),  are shown 
in the figure as dashed lines.} \label{fig:p_large_rho}
\end{figure}

\begin{figure}
\centerline{\scalebox{0.45}{
\includegraphics{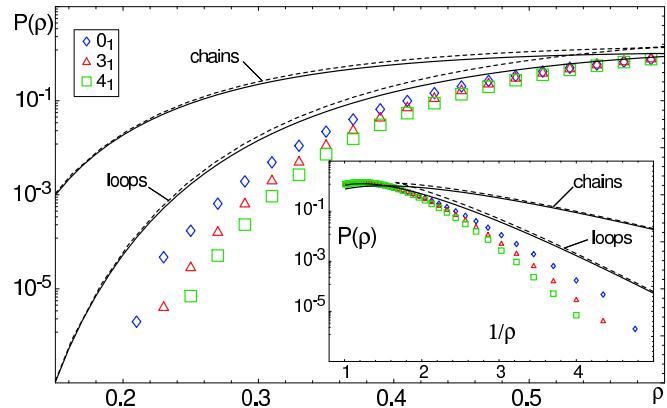}}}
\caption{The probability density plot for
chains\protect\cite{Fixman} (line), all loops (another line), and
loops with certain knots ($0_1$ - $\diamond$, $3_1$ - $\Delta$,
$4_1$ -$\Box$) in the range of small $\rho < 1$. Distributions are
presented in terms of the scaling variable $\rho = R_{g}^{2} /
\langle R_{g}^{2} \rangle$.  The asymptotics calculated in \ref{sec:Rg_distribution}, 
equations (\ref{eq:asymp_chain}) and (\ref{eq:asymp_loops}),  are shown 
in the figure as dashed lines.  \textbf{Inset:} Semi-log probability
density plot (or linear entropy plot) at small $\rho$ against
$1/\rho$. } \label{fig:p_small_rho}
\end{figure}

\begin{figure}
\centerline{\scalebox{0.45}{
\includegraphics{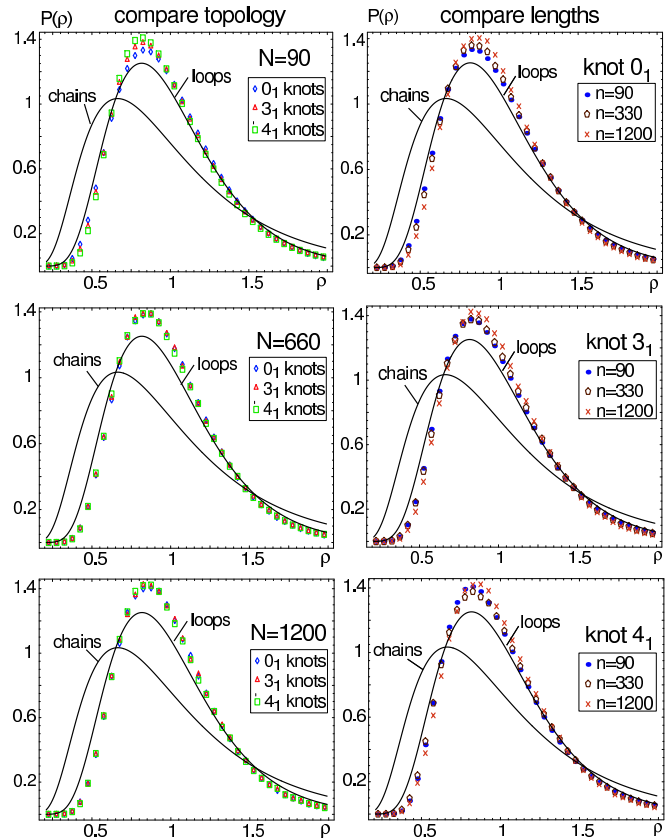}}}
\caption{The probability distributions, $P( \rho )$ for several
different lengths, $N$.  \textbf{Left Column:} Collapse of several
different topologies to one curve at large $N$, (compare $N=660$
or $1200$ to $N=90$), implies that one master curve for
under-knotted loops exists, and that it is visible for $0_1$,
$3_1$, and $4_1$ knots at $N \ge 660$. \textbf{Right Column:}
Curves for these simple topologies, as they differ in length, are
certainly more similar to each other than they are to the average
of all loops. Movement of the curves as $N$ changes is not yet
understood. }\label{fig:compare_curves}
\end{figure}

Our main findings are summarized in figures (\ref{fig:p_large_rho}),
(\ref{fig:p_small_rho}), and (\ref{fig:compare_curves}), where we present probability
distributions $P(\rho)$ for the trivial knots $0_{1}$
($\diamond$), trefoils $3_1$ ($\Delta$), and $4_1$ knots ($\Box$).
In the same figures we plot also for comparison the analytically
computed probability distributions for linear chains and for all
loops.  For linear chains, the necessary distribution $P_{\rm
chain} (\rho)$ was found by Fixman a long time ago\cite{Fixman};
as described in \ref{sec:Rg_distribution}, we were able to derive
a similar expression for the probability distribution over all
loops, irrespective of topology.  To avoid overloading the
figures, we do not show the corresponding data points obtained for
linear chains and for all loops, but they all sit essentially on
top of the theoretical curves (confirming once again the
ergodicity of our loop generation routine).

Comparing the shapes of probability distributions for all loops
and those with identified quenched topology, we notice that the
latter distributions are somewhat more narrow.  We emphasize, that
although the effect looks small for the eye, it is well above the
error bars of our measurements.  This means simple knots are less
likely to swell much above their average size than other knots,
and they are also less likely to shrink far below their average,
again compared to other knots. Figures \ref{fig:p_large_rho} and
\ref{fig:p_small_rho} show this stiffness in both large and small
limits of $\rho$.  In the region $\rho<1.25$ the general notion that
entropic stiffness goes with topological complexity seems to hold true,
i.e. more complex knots are more difficult to stretch or compress
than arbitrary loops of the same number of segments.  That the opposite
of this seems to be true in the large $\rho$ region is a subtlety
not yet fully understood.  In any case, topology blind loops are
by definition always more flexible than topology specific loops.

The small $\rho$ limit is of particular interest given its
relation to all problems involving collapsed polymers, such as
proteins. A closer view of the small $R_g$ region of the
probability distribution is presented in the Figure
\ref{fig:p_small_rho}. There, the probability distributions are
plotted in the semi-$log$ scale against $\rho$ and, in the inset,
against $1/\rho$.   This can be also understood as the plot of
"confinement" entropy, which corresponds to the squeezing the
polymer to within certain (small) radius. The reason why we plot
the data against $1/\rho$ is because both $P_{\rm chain} (\rho)$
and $P_{\rm loops}( \rho )$ at small $\rho$ have asymptotics $\sim
\exp \left( - {\rm const} / \rho \right)$ (see formulae
(\ref{eq:asymp_chain}) and (\ref{eq:asymp_loops})), which
corresponds to confinement entropy $\sim 1/\rho$, and which can be
established by a simple scaling argument, as described, e.g., in
the book\cite{AG_Red} (page 42). This $1/\rho$ behavior is seen
clearly in Figure (\ref{fig:p_small_rho}). Furthermore, we see
indeed that compressing any specific knot, trivial or otherwise,
is significantly more difficult than compressing a phantom loop.
Analytical expression of entropy for knots is not known, thus far
only the $R_{g}^{-3} \sim \rho^{-3/2}$ scaling at small $\rho$ has
been conjectured\cite{crumpled}. Although our data is
qualitatively consistent with this prediction in terms of the
direction of the trend, more data is needed for quantitative
conclusion.

More detailed comparison of probability distributions for
different knots and different $N$ are presented in the Figure
\ref{fig:compare_curves}.  This figure shows a number of different
probability curves under different conditions.  The left column of
this figure compares the topology of different objects while
holding the length of the objects constant.  The right column of
the same figure shows comparisons of different lengths of the same
topology.  Significant in Figure (\ref{fig:compare_curves}) is the
suggestion that probability distributions for different knots
become very similar if not identical with increasing $N$.  Indeed
in the left column in Figure \ref{fig:compare_curves} it is
difficult to see the difference between the distributions for the
three distinct topologies for $N=1200$ or even for $N=660$.  One
way to understand this effect is to consider the notion of knot
localization\cite{localization_1,localization_2,localization_3}.
The idea is that every strongly under-knotted loop at large $N$
places its knot in some small fraction of its length, thus looking
like  a trivial knot, with a small bump where the appropriate
crossings reside.  The collapse of $P_{{\cal K}}(\rho)$ for
different, simple knot types, ${\cal K}$, to one curve at large
$N$ is consistent with this concept of localization.  At the same
time, Figure \ref{fig:compare_curves} suggests that probability
distribution $P_{{\cal K}}(\rho)$ for each knot keeps evolving
with $N$ changing over the cross-over region at $N \sim N_0$.

\section{Concluding remarks}
To summarize, in this paper we presented computational results on
knots in zero thickness loops of $N$ rigid segments of equal
length $\ell$.  To the accuracy of our measurements, our data are
consistent with the idea that mean square gyration radius averaged
over the loops which are topologically equivalent to trivial knots
is larger than the similar quantity averaged over all loops
irrespective of topology.  The extent of this additional swelling
appears similar to the swelling of self-avoiding walks compared to
Gaussian random walks.  Swelling is characteristic not only of
trivial knots, but in general for under-knotted loops, in the
sense that a topologically quenched loop swells if its knot state
would have simplified upon annealing of its topology.  We have
examined not only averaged gyration radius, but also its
probability distribution.  We found that topologically
under-knotted loops are relatively unlikely to deviate far from
their average sizes, either to smaller or to larger sizes.  We have
also found indication that the probability distribution of the
gyration radius of simple knots becomes universal for all under-knotted loops when
their length exceeds certain threshold.  Importantly, our data
confirm the existence of a cross-over at $N$ of the order of
$N_0$, the characteristic length of random knotting: it is only at
$N > N_0$ that there is analogy between under-knotted loops and
self-avoiding walks, at $N < N_0$ topological constraints have
only a marginal effect on the trivial knots.

How far does the analogy go between self-avoiding polymers and
topologically constrained ones?  We were unable to confirm this
analogy beyond simple scaling; it is unclear whether the $\langle
R_g^2 \rangle $ dependence on $N$ approaches its scaling form
$N^{2 \nu}$ in the same manner as it happens for self-avoiding
walks.  It is worth emphasizing that there is a field theoretic
formulation for the self-avoiding walks\cite{deGennes_th}, but there
is nothing of this sort for knots.  In our opinion, it remains an
exciting challenge to find a solid understanding of the connection
between fluctuation properties of the loop and its topology.

\section*{Acknowledgments}

We thank A.Vologodskii for sharing with us his unpublished
data\cite{VOLOGODSKII_private} and T.Deguchi for useful
discussions.  This work was supported in part by the MRSEC Program
of the National Science Foundation under Award Number DMR-0212302.

\appendix

\section{Loop generation}\label{sec:Loop_Generation}

We generated loops of the length $N$ divisible by $3$ using the
following method.   To produce one loop, we generated $N/3$
randomly oriented equilateral triangles of perimeter $ 3 \ell$. We
consider each triangle a triplet of head-to-tail connected
vectors. Collecting all $N$ vectors from $N/3$ triangles, we
re-shuffled them, and connected them all together, again in the
head-to-tail manner, thus obtaining the desired closed loop.

\begin{figure}
\centerline{\scalebox{0.45}{
\includegraphics{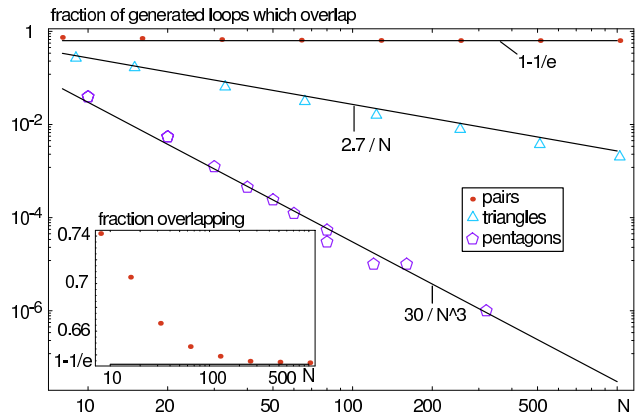}}}
\caption{The fraction of generated loops which overlapped within
the resolution of computational accuracy.  If a set of $N/s$ polygons, where
polygons have $s$ sides, is used to generate a walk of length $N$, the
fraction of generated loops, $\sim N^{2-s}$ will overlap exactly.  This behavior is seen in the image.}
\label{fig:fraction_missed}
\end{figure}

A similar simpler method applicable for even $N$ and re-shuffling
vectors obtained from zero sum pairs often yields the loops with
overlapping nodes.  This happens when the re-shuffling results in
the succession of some $2m < N$ vectors belonging to exactly $m$
pairs and thus forming the zero sum (i.e., closed) sub-loop.  The
probability of such an event is of order unity, because the
probability for the two vectors from the same pair to be next to
each other after the re-shuffling scales as $1/N$, and there are
$\sim N$ such pairs; more accurate calculation\cite{combinatorics}
shows that this probability approaches $1-1/e$ as $N \to \infty$.

For the triangles, the problem is not in any way as severe, because the
probability for the three vectors of the triplet to be next to
each other scales as $1/N^2$, while the number of triangles is
still $\sim N$, so the overlapping loops are rare as $1/N$ (and
the probability to have two, or, in general, $m$ triplets to
occupy completely the $3m$ stretch of the re-shuffled sequence
does not change the $1/N$ estimate).

\begin{figure}
\centerline{\scalebox{0.45}{
\includegraphics{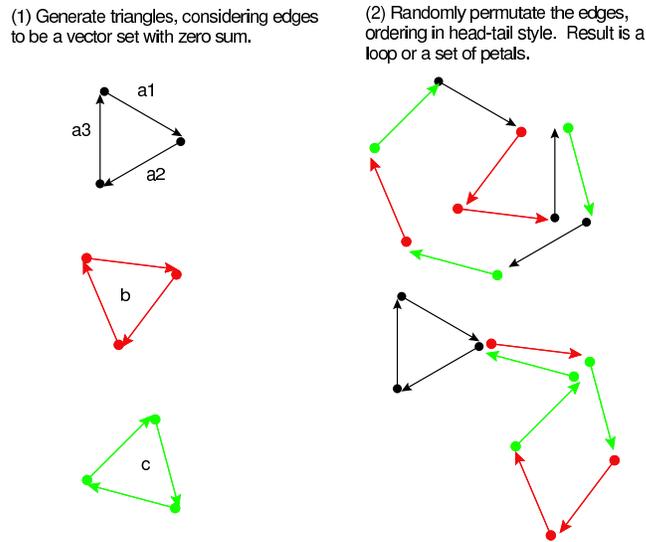}}}
\caption{Our generation routine can produce errant objects which are
not loops. } \label{fig:petals}
\end{figure}

Our test measurements of the fraction of loops overlapping generated
with pairs, triplets, and pentagons of vectors (squares are $2$ pairs), shown in figure
(\ref{fig:fraction_missed}), agree with this understanding.
We see in this figure that the fraction overlapping at a certain
$N$, when generated in polygons of $s$ edges scales like
$N^{2-s}$.  We chose to generate with triplets to avoid the
constant overlap implied by pairs, as well as avoiding the
correlation implicit with larger sets of objects.  Although
generated with our method, these loops are not members of the set
analyzed as they are not single stranded loops devoid of
self-intersections, but rather a different physical class of
objects with "petals."  The simple example of $N=9$, see figure
(\ref{fig:petals}), illustrates this.  Suppose that the three
triangles generated have segment vectors,
$(\vec{a_1},\vec{a_2},\vec{a_3}),$
$(\vec{b_1},\vec{b_2},\vec{b_3})$ and
$(\vec{c_1},\vec{c_2},\vec{c_3})$. By definition, each set of
vectors within a triangle sums to zero, for example,
$\vec{c_1}+\vec{c_2}+\vec{c_3}=0$. A walk is then created by a
random permutation of all of the segment vectors, for example, $
(\vec{a_3},\vec{c_2},\vec{a_1}, \vec{b_1},\vec{a_2}, \vec{b_3},
\vec{c_1},\vec{b_2},\vec{c_3})$. The problem of overlapping,
described above, occurs whenever the elements of one or more
complete triangles occur within a continuous section of the
permutation vector.  This subsection forms a complete loop, as
does the rest of the chain and instead of a single loop, one has a
diagram which looks something like a flower with multiple petals
coming off of a center axis or set of axes.

In practice, there was also a totally different problem.  At large
$N$, our knot identification routine was sometimes failing because
of the perceived triple crossing on the projection.  A simple
rotation by random Euler angles resolved this projection problem
in all cases.

\section{Probability distribution of all loops}\label{sec:Rg_distribution}

In this Appendix we address the problem which, of its own, does
not belong to the subject of knots.  Namely, we consider a phantom
loop, which can freely pass through itself, and determine the
probability distribution of its gyration radius.  In other,
equivalent, words we consider the distribution of sizes over the
ensemble of all possible loops of the given number of segments,
$N$, irrespective of their topology.  Our approach here closely
follows that of the work\cite{Fixman} by Fixman, where he
determined probability distribution for the gyration radius of the
linear chains.  To make our work self-contained, we reproduce
below the main steps of Fixman derivation along with our results
for phantom loops.

To begin with, we simplify the problem by transforming it from the
gyration radius of a chain or a loop with $N$ rigid segments of
fixed length $\ell$ to the similar problem with a smaller number
of Gaussian distributed segments.  To achieve this, we group $N$
segments in $n$ blobs of $N/n$ segments each.  We denote as $b
{\vec \eta}_{k}$ the end-to-end vector of each blob labeled $k$,
where $b^2 = (N/n) \ell^2$.  Note that mean squared gyration
radius, which is well known for both chains and loops\cite{AG_Red},
can be expressed in terms of either $N$
and $\ell$ or $n$ and $b$: $\langle R_{g}^{2} \rangle_{\rm chain}
= N \ell^{2}/6 = n b^{2}/6$ for chains and $\langle R_{g}^{2}
\rangle_{\rm loop} = N \ell^{2}/12 = n b^{2}/12$ for loops.

If $N/n \gg 1$, then probability distribution for the unitless
vector ${\vec \eta}$ is Gaussian, with zero mean and unit
variance: $ g({\vec \eta}) = \left(3 / 2 \pi  \right)^{3/2} \exp
\left( - 3 \eta^2 / 2 \right) $.  If, at the same time, $n \gg
1$, then computing the gyration radius (\ref{eq:gyration}) we can
replace each blob with the concentrated mass $N/n$ sitting, say,
at the beginning segment of this blob. Then, formula
(\ref{eq:gyration}) can be transformed to have just $n$ (instead
of $N$) points, where now ${\vec r}_{ij} = b \sum_{k=i}^{j} {\vec
\eta}_{k}$.  Accordingly, the gyration radius can be expressed as
a quadratic form of the vectors ${\vec {eta}}$.  It is convenient to
write it in the form
\begin{equation} \rho \equiv \frac{R_{g}^{2}}{\langle R_{g}^{2} \rangle}
= A \sum_{k,m=1}^{n} G(k,m) \ {\vec \eta}_{k} \cdot {\vec
\eta}_{m} \ , \label{eq:rho_quadratic_form} \end{equation}
where coefficient $A$ is different for chains and loops and can be
determined at the end to ensure the correct average ($\langle \rho
\rangle =1$), and where kernel $G(k,m)$ is as follows:
\begin{equation} G(k,m) = \frac{k}{n^2} H(m-k) + \frac{m}{n^2} H(k-m)
- \frac{km}{n^3} \ ; \ H(x) = \left\{ \begin{array}{lcr} 1 & {\rm
for} & x >0 \\ 1/2 & {\rm for} & x =0 \\0 & {\rm for} & x <0
\end{array} \right. \ .  \end{equation}

We now note that the probability of the chain conformation
specified by blob end-to-end vectors $b{\vec \eta}_{1}, b{\vec
\eta}_{2}, \ldots , b{\vec \eta}_{n}$ is given by
\begin{equation} Z_{\rm chain} ( \left\{ {\vec \eta} \right\}) =
\prod_{k=1}^{n-1} g({\vec \eta}_{k}) \ .
\label{eq:Z_for_chains}\end{equation}
Similar probability for the loop reads
\begin{equation} Z_{\rm loop} ( \left\{ {\vec \eta} \right\}) =
\prod_{k=1}^{n} g({\vec \eta}_{k}) \times \delta \left(
\sum_{k=1}^{n} {\vec \eta}_{k} \right) \times \left( \frac{2 \pi n
}{3} \right)^{3/2} \ . \label{eq:Z_for_loops}
\end{equation}
Compared to the distribution for the chains, we have here one more
factor $g$, describing the connection between chain head and tail,
making the loop; we have $\delta$-function ensuring loop closing;
and we have also the normalization factor.

Now, in order to compute probability distribution of $\rho$, we
introduce the characteristic function
\begin{equation} K(s) = \langle e^{\imath \rho s} \rangle = \int e^{\imath \rho
s} Z\left( \left\{ {\vec \eta } \right\} \right) d \left\{ {\vec
\eta } \right\} \ ,
\end{equation}
where $Z$ is either $Z_{\rm chain}$ or $Z_{\rm loop}$.  Looking at
the expressions for $Z$, (\ref{eq:Z_for_chains}) or
(\ref{eq:Z_for_loops}), and for $\rho$,
(\ref{eq:rho_quadratic_form}), we see that the three Cartesian
components of vectors ${\vec \eta}$ decouple.  Taking advantage of
this decoupling, we can write $K(s) = \left[ f(s) \right]^{3}$,
where
\begin{eqnarray} f_{\rm chain}(s) & = & \left( \frac{3}{2 \pi} \right)^{(n-1)/2}
\int \exp \left[ - \frac{3}{2} \sum_{k=1}^{n-1} \eta_{k}^{2} +
\right. \nonumber \\ & + & \left. \imath s A \sum_{k,m=1}^{n-1}
G(k,m) \eta_{k} \eta_{m} \right] d \eta_{1} d \eta_{2} \ldots d
\eta_{n-1}
\end{eqnarray}
for chains, and
\begin{eqnarray} f_{\rm loop}(s) & = & \left( \frac{3}{2 \pi}
\right)^{n/2} \left( \frac{n}{6 \pi} \right)^{1/2} \int \exp
\left[ \imath p \sum_{k=1}^{n} \eta_{k}- \frac{3}{2}
\sum_{k=1}^{n-1} \eta_{k}^{2} + \right. \nonumber \\ & + & \left.
\imath s A \sum_{k,m=1}^{n-1} G(k,m) \eta_{k} \eta_{m}   \right] d
\eta_{1} d \eta_{2} \ldots d \eta_{n} dp
\end{eqnarray}
for loops.  In the later case, we have used the integral
representation of the $\delta$-function, thus the extra
integration over $p$.  These Gaussian integrals are easy to
evaluate, because the matrix $G(k,m)$ is diagonalized, (we have omitted
details\cite{Fixman}), by the unitary matrix $C(k,m) =
\sqrt{2/n} \sin \left( \pi k m /n \right)$, revealing the
eigenvalues of the $G$ matrix, $1/k^2 \pi^2$ with all integer $k$
from $1$ to $n$.  Upon some algebra, we obtain for chains
\begin{eqnarray} f_{\rm chain}(s) & = &
 \prod_{k=1}^{n-1} \left( 1 - \imath \frac{2 sA}{3 k^2 \pi^2
}\right)^{-1/2} \simeq \nonumber \\ & \simeq & \left[\underbrace{
\prod_{k=1}^{\infty} \left( 1 - \frac{z^2}{k^2 \pi^2}
\right)}_{\sin z/z} \right]^{-1/2} = \left( \frac{z}{\sin z}
\right)^{1/2} \ ,
\end{eqnarray}
where $z^2 = 2 \imath A s/3$.  Similar manipulations for loops
involve an extra integral over $p$:
\begin{eqnarray} f_{\rm loop}(s) & = & \left( \frac{z}{\sin z}
\right)^{1/2} \left( \frac{n}{6 \pi} \right)^{1/2}
\int_{-\infty}^{\infty} \exp \left[-p^2 \frac{n}{3} \sum_{k=1}^{n}
\frac{(1 -(-1)^{k})^{2}}{\pi^2 k^2 - z^2} \right] dp \simeq
\nonumber
\\ & \simeq &  \left( \frac{z}{\sin z} \right)^{1/2}
\left[ 8 \underbrace{\sum_{m=0}^{\infty} \frac{1}{\pi^2 (2 m +
1)^2 - z^2}}_{\tan (z/2) / 4 z} \right]^{-1/2} = \frac{z/2}{\sin
(z/2)} \ ,
\end{eqnarray}
where again $z^2 = 2 \imath A s /3$.  Finally, we choose
coefficient $A$ based on the condition $\langle \rho \rangle =1$,
or $K^{\prime}(s)_{s=0}=\imath$.  This yields $A=6$ for chains and
$A=12$ for loops.  Therefore, we finally get
\begin{equation} K_{\rm chain}(s)=\left( \sin z /z \right)^{-3/2}
\ , \ \ \ z^{2}=4 \imath s \ , \end{equation}
(the result due to Fixman\cite{Fixman}), and
\begin{equation} K_{\rm loops}(s) = \left( 2 \sin(z/2) / z
\right)^{-3} \ , \ \ \ z^{2} = 8 \imath s \ . \end{equation}

Knowing $K(s)$, finding the probability distribution $P(\rho)$ is
the matter of inverse Fourier transform.  Numerical inversion of
Fourier transforms yield the curves presented in the Figures
\ref{fig:p_large_rho}, \ref{fig:p_small_rho}, and
\ref{fig:compare_curves}.

Analytically, asymptotic expressions can be found for both small
and large $\rho$.  For chains, Fixman\cite{Fixman} found
\begin{equation} P_{\rm chain} (\rho) \simeq \left\{\begin{array}{lcr}
\frac{\pi^{5/2}e^{3/2}}{6} \rho^{1/2} e^{-\rho \pi^2 / 4} &{\rm for}& \rho \gg 1 \\
9 \sqrt{\frac{6}{\pi}} \rho^{-3} e^{-9/(4 \rho)}& {\rm for} & \rho
\ll 1 \end{array} \right. \ . \label{eq:asymp_chain}
\end{equation}
Similar expressions for loops read
\begin{equation} P_{\rm loop} (\rho) \simeq \left\{\begin{array}{lcr}
\frac{\pi^{6}}{2} \rho^{2} e^{-\rho \pi^2 / 2} &{\rm for}& \rho \gg 1 \\
324 \sqrt{\frac{2}{\pi}} \rho^{-9/2} e^{-9/(2 \rho)}& {\rm for} &
\rho \ll 1 \end{array} \right. \ . \label{eq:asymp_loops}
\end{equation}
To obtain these results, it is convenient to re-write the inverse
Fourier transform:
\begin{equation}  P_{\rm loop} = \frac{1}{2 \pi} \int K(s)
e^{- \imath s \rho} ds = \frac{1}{2 \pi \imath} \int_{V}
\frac{\zeta^4}{\sin^3 \zeta} e^{-\rho \zeta^2/2} d \zeta \ ,
\end{equation}
where in the latter integral $\zeta = z/2$ and integration contour
$V$ in complex $\zeta$-plane is V-shaped, runs from infinity along
the line with argument $3 \pi /4$ to infinity along the line with
argument $\pi/4$.  In this form, it is conveniently seen that
$P_{\rm loop} (\rho) =0$ at $\rho < 0$, as it must be, since
$\rho$ is a positive quantity.  Furthermore, deforming the
integration contour, we can establish that at $\rho \ll 1$ the
integral is dominated by the saddle at $\zeta \simeq 3 \imath /
\rho$, while at $\rho \gg 1$ it is dominated by the residue at the
third order pole in $\zeta = \pi$, yielding the results
(\ref{eq:asymp_loops}).

On a more physical note, it is important to realize that the
exponential terms in equations (\ref{eq:asymp_chain}) and
(\ref{eq:asymp_loops}) at small $\rho$ are identical if written in
terms of $R_{g}$, $N$ and $\ell$ instead of $\rho$.  Indeed, the
leading term of the corresponding entropy (which is $-\ln P$) is
equal to $9 N \ell^2 / 24 R_g^2$ for both chains and loops.  Apart
from the coefficient of $9/24$, the scaling form of this result
can be understood based on a simple argument considering
confinement of either a chain or a loop in a cavity of the size $R
\ll \ell \sqrt{N}$ (see, for instance, book \cite{AG_Red}, formula
(7.2)).

On the other hand, at large $\rho$ chain entropy is $3 (\pi
R_g)^2/2 N \ell^2$, while loop entropy is four times larger, it is
$6 (\pi R_g)^2/ N \ell^2$.  This can be understood as follows. For
the chain, remembering that entropy of the state with end-to-end
distance $L$ is $3 L^2/2 N \ell^2$, Fixman noted \cite{Fixman}
that large $R_g$ conformations are dominated by the semi-circular
shapes with $L = \pi R_g$.  The loop obviously represents two such
pieces, so loop entropy is twice entropy of the half-chain: $6
(\pi R_g)^2/ N \ell^2 = 2 \times \left(3 (\pi R_g)^2/2 (N/2)
\ell^2 \right)$.

\end{document}